\documentclass[prb,twocolumn,floatfix,amsmath,amssymb,showpacs]{revtex4}
\usepackage{graphicx}
\usepackage{dcolumn}
\usepackage{bm}

\begin{document}


\title{Muon spin relaxation and hyperfine-enhanced $\bm{^{141}}$Pr nuclear spin dynamics in Pr(Os,Ru)$\bm{_{4}}$Sb$\bm{_{12}}$ and (Pr,La)Os$\bm{_4}$Sb$\bm{_{12}}$}

\author{Lei Shu}
\author{D.~E. MacLaughlin}
\affiliation{Department of Physics and Astronomy, University of California, Riverside, California 92521}
\author{Y. Aoki}
\author{Y. Tunashima}
\author{Y. Yonezawa}
\author{S. Sanada}
\author{D. Kikuchi}
\author{H. Sato}
\affiliation{Department of Physics, Tokyo Metropolitan University, Tokyo 192-0397, Japan}
\author{R.~H. Heffner}
\altaffiliation{Present address: Los Alamos National Laboratory, Los Alamos, New Mexico 87545.}
\author{W. Higemoto}
\author{K. Ohishi}
\author{T.~U. Ito}
\affiliation{Advanced Science Research Center, Japan Atomic Energy Agency, Tokai-Mura, Ibaraki 319-1195, Japan}
\author{O.~O. Bernal}
\affiliation{Department of Physics and Astronomy, California State University, Los Angeles, California 90032}
\author{A.~D. Hillier}
\affiliation{ISIS Facility, Rutherford Appleton Laboratory, Chilton, Didcot, Oxfordshire, OX11 0QX, United Kingdom}
\author{R. Kadono}
\author{A. Koda}
\affiliation{Meson Science Laboratory, KEK, Tsukuba, Ibaraki 305-0801, Japan}
\author{K. Ishida}
\affiliation{ Department of Physics, Graduate School of Science, Kyoto University, Kyoto 606-8502, Japan}
\author{H. Sugawara}
\affiliation{Faculty of the Integrated Arts and Sciences, The University of Tokushima, Tokushima 770-8502, Japan}
\author{N.~A. Frederick}
\author{W.~M. Yuhasz}
\author{T.~A. Sayles}
\author{T.~Yanagisawa}
\author{M.~B. Maple}
\affiliation{Department of Physics and Institute for Pure and Applied Physical Sciences, University of California, San Diego, La Jolla, California 92093}

\date{\today}

\begin{abstract}
Zero- and longitudinal-field muon spin relaxation ($\mu$SR) experiments have been carried out in the alloy series~Pr(Os$_{1-x}$Ru$_x$)$_4$Sb$_{12}$ and Pr$_{1-y}$La$_y$Os$_4$Sb$_{12}$ to elucidate the anomalous dynamic muon spin relaxation observed in these materials. The damping rate~$\Lambda$ associated with this relaxation varies with temperature, applied magnetic field, and dopant concentrations~$x$ and $y$ in a manner consistent with the ``hyperfine enhancement'' of $^{141}$Pr nuclear spins first discussed by Bleaney in 1973. This mechanism arises from Van Vleck-like admixture of magnetic Pr$^{3+}$ crystalline-electric-field-split excited states into the nonmagnetic singlet ground state by the nuclear hyperfine coupling, thereby increasing the strengths of spin-spin interactions between $^{141}$Pr and muon spins and within the $^{141}$Pr spin system. We find qualitative agreement with this scenario, and conclude that electronic spin fluctuations are not directly involved in the dynamic muon spin relaxation.
\end{abstract}


\maketitle

\section{\label{sect:Intro}Introduction}
The filled skutterudite compound~PrOs$_{4}$Sb$_{12}$ is the first praseodymium-based heavy-fermion superconductor to be discovered,\cite{BFHZ02} and is one of the few $f$-electron heavy-fermion compounds in which a non-Kramers ion exhibits a nonmagnetic crystalline-electric-field (CEF)-split ground state. Both the normal and superconducting states of PrOs$_{4}$Sb$_{12}$ are unusual: The $\Gamma_1$ singlet Pr$^{3+}$ ground state is separated from a $\Gamma_{4}^{(2)}$ first excited state (tetrahedral notation\cite{THY01}) by a remarkably small splitting~$\Delta E_{\mathrm{CEF}}/k_B \approx 7$~K,\cite{ANOS02,MHZF02,KINM03,GOBM04} leading to strong CEF effects at low temperatures. There is no sign of magnetic ordering. The Sommerfeld coefficient~$\gamma$ is difficult to measure in the presence of the CEF Schottky anomaly in the low-temperature specific heat, but is estimated to lie between 500 and $750~{\mathrm{mJ~mol^{-1}~K^{-2}}}$. The Pr$^{3+}$ ions are enclosed in an icosahedral cage of Sb atoms that is considerably larger than the ionic size, and large-amplitude Einstein-like $4f$-ion phonon modes (``rattling'' modes) are observed. Below the superconducting transition temperature~$T_c = 1.85$~K an unconventional superconducting phase is found,\cite{ATSK07} with evidence for multiple phases, time reversal symmetry breaking,\cite{ATKS03} and extreme multiband behavior. \cite{SBMF05,SBMB06} Dispersive antiferroquadrupolar excitons and an unusual low-temperature high-field phase with antiferroquadrupolar order are observed. \cite{KINM03}  The mechanism or mechanisms for heavy-fermion behavior and Cooper pairing in this compound remain controversial; rattling modes\cite{GNSY04} and Pr$^{3+}$ quadrupole fluctuations\cite{MKH03} have been proposed.

Muon spin relaxation ($\mu$SR),\cite{Sche85,Brew94} like other magnetic resonance techniques, probes magnetism and electronic structure in solids on the microscopic (atomic) size scale. In $\mu$SR experiments spin-polarized positive muons ($\mu^+$) are implanted into the material of interest and stop at interstitial sites. During its lifetime each muon spin precesses in the magnetic field at its site and eventually decays ($\mu^+ \rightarrow e^+ + \nu_e + \overline{\nu}_\mu$); the direction of the emitted positron's momentum is correlated with the muon spin direction at the moment of decay. A large number of such events determines the time development (relaxation) of the ensemble-average muon spin polarization. Several $\mu$SR studies of Pr-based filled skutterudites have been reported. \cite{MSHB02,ATKS03,AHSO05,SHKO05,AHPG05,HFKW06,HAOI07}

The magnetic environment of the muon creates a local field~$\mathbf{H}_{\mathrm{loc}}$ at the muon site that causes muon spin relaxation. Relaxation mechanisms can be divided into two classes, depending on the behavior of $\mathbf{H}_{\mathrm{loc}}$:
\begin{itemize}

\item static (or quasistatic\cite{quasistatic}) relaxation, due to an inhomogeneous static distribution of $\mathbf{H}_{\mathrm{loc}}$ that causes a spread of muon Larmor frequencies and consequent loss of muon spin phase coherence;

\item dynamic relaxation (often called spin-lattice relaxation in the NMR literature), due to thermal fluctuations of $\mathbf{H}_{\mathrm{loc}}$ that induce transitions between muon spin levels and equilibrate the muon spin populations. The equilibrium muon spin polarization is negligible compared to the initial polarization ($\sim$100\%). The dynamic component of $\mathbf{H}_{\mathrm{loc}}$ usually arises from electronic spin fluctuations; we shall see, however, that nuclear spin fluctuations can also be involved.

\end{itemize}
In longitudinal-field $\mu$SR (LF-$\mu$SR), which includes zero-field $\mu$SR (ZF-$\mu$SR) as a special case, a magnetic field~$H_L$ is applied parallel to the initial muon spin polarization. The dependence of the muon spin relaxation on $H_L$ helps to separate the static and dynamic contributions to the relaxation. \cite{KuTo67,HUIN79}

The ZF-$\mu$SR spin relaxation function in PrOs$_{4}$Sb$_{12}$ at low temperatures\cite{ATKS03} could be fit by the product of an exponential damping factor~$\exp(-\Lambda t)$ and the Kubo-Toyabe (\mbox{K-T}) functional form\cite{KuTo67,HUIN79} expected for a Gaussian quasistatic field distribution. This behavior was attributed to a two-component form of $\mathbf{H}_{\mathrm{loc}}$: a static component, responsible for the K-T relaxation, and a second component, responsible for the exponential damping, which was determined by LF-$\mu$SR measurements to be due to dynamic fluctuations. \cite{ATKS03} The origin of this dynamic component was not clear, although $4f$-electron dynamics associated with the small CEF splitting were noted as a possible mechanism. In the normal state above $T_c$ the quasistatic field distribution was attributed to dipolar interactions between muon spins and neighboring (principally $^{121}$Sb and $^{123}$Sb) nuclear magnetic moments. An increase in the quasistatic relaxation rate was observed below $T_c$, and interpreted as evidence for time-reversal symmetry breaking in the superconducting state of PrOs$_{4}$Sb$_{12}$. \cite{ATKS03} This effect has also been studied in Pr(Os$_{1-x}$Ru$_x$)$_{4}$Sb$_{12}$ and Pr$_{1-y}$La$_y$Os$_{4}$Sb$_{12}$ alloys,\cite{SHAF07,Shu07} which are superconducting for all $x$ and $y$ with $T_c$ of the order of 1~K. \cite{FDHB04,RKA06}

In this article we report results of ZF- and LF-$\mu$SR experiments in the Pr(Os$_{1-x}$Ru$_x$)$_{4}$Sb$_{12}$ and Pr$_{1-y}$La$_y$Os$_{4}$Sb$_{12}$ alloy systems,\cite{otherskuttmuSR} which have been undertaken to elucidate the anomalous exponential damping. We confirm that the damping is indeed dynamic in nature, and argue that it is due to enhancement of $^{141}$Pr nuclear magnetism via intra-atomic hyperfine coupling to the Pr$^{3+}$ $4f$ electrons. \cite{Blea73} This coupling induces a Van Vleck-like admixture of low-lying Pr$^{3+}$ magnetic CEF-split excited states into the singlet ground state, thereby increasing the effective $^{141}$Pr nuclear moment. Both the $^{141}$Pr-$\mu^+$ and the $^{141}$Pr-$^{141}$Pr dipolar interactions are increased (by factors of $\sim$20 and $\sim$400, respectively) by this ``hyperfine enhancement'' mechanism, which also increases indirect $^{141}$Pr-$^{141}$Pr interactions mediated by exchange coupling between Pr$^{3+}$ ions. \cite{Blea73} Hyperfine-enhanced $^{141}$Pr nuclear moments have a significant effect on muon spin relaxation: the mechanism accounts for the observed damping of the muon spin relaxation, and is qualitatively consistent with the behavior of the damping rate~$\Lambda$ with enhancement strength and Pr-ion concentration dependence (for Pr$_{1-y}$La$_y$Os$_{4}$Sb$_{12}$ alloys) across both alloy series. Exchange-mediated interactions appear to dominate the $^{141}$Pr spin dynamics. The origin of the dynamical muon spin relaxation has been clarified, and we conclude that {\em electronic\/} spin fluctuations (except those associated with the hyperfine-enhancement mechanism) are not directly involved in the anomalous muon spin relaxation.

The article is organized as follows. After a brief description of the experiments in Sec.~\ref{sec:exp}, we report our experimental results for Pr(Os$_{1-x}$Ru$_{x}$)$_{4}$Sb$_{12}$ and Pr$_{1-y}$La$_{y}$Os$_{4}$Sb$_{12}$ in Sec.~\ref{sec:results}. We consider the dependence of the muon exponential damping rate~$\Lambda$ on $H_L$, temperature, and dopant concentrations $x$ and $y$, and their implications for the mechanism for the dynamic muon spin relaxation in Sec.~\ref{sec:discussion}. 
Our conclusions are summarized in Sec.~\ref{sec:concl}.

\section{Experiments} \label{sec:exp}
ZF- and LF-$\mu$SR experiments were carried out in
Pr(Os$_{1-x}$Ru$_x$)$_{4}$Sb$_{12}$, $x = 0.05$, 0.1, 0.2, 0.6,
and 1.0, and (Pr$_{1-y}$La$_y$)Os$_{4}$Sb$_{12}$, $y = 0.2$, 0.4,
0.6, and 0.8. ZF- and LF-$\mu$SR experiments were carried out on
powdered samples at the Meson Science Laboratory, KEK, Tsukuba,
Japan, and the MuSR spectrometer at the ISIS neutron and muon
facility, Rutherford Appleton Laboratory, Chilton, U.K\@.
LF-$\mu$SR experiments were carried out at the M15 beam line at
TRIUMF, Vancouver, Canada, on mosaics of oriented $\sim$1 mm$^3$
crystals, prepared by the Sb-flux method,\cite{BSFS01} with
$\langle 100\rangle$ directions parallel to the applied field.

$^3$He-$^4$He dilution cryostats were used to obtain low temperatures. The mosaic crystals were mounted on a thin GaAs backing, which at low temperatures rapidly depolarizes muons and minimizes any spurious signal from muons that do not stop in the sample. Standard time-differential $\mu$SR asymmetry data\cite{Sche85,Brew94} were taken in the normal and superconducting states at temperatures in the neighborhood of $T_c$ for $H_L$ between 0 and 125~Oe.

\section{Experimental Results} \label{sec:results}
\subsection{\label{sect:ZF}Zero-field muon spin relaxation} \label{sec:ZF}
\begin{figure}[ht]
 \begin{center}
 \includegraphics*[width=0.45\textwidth]{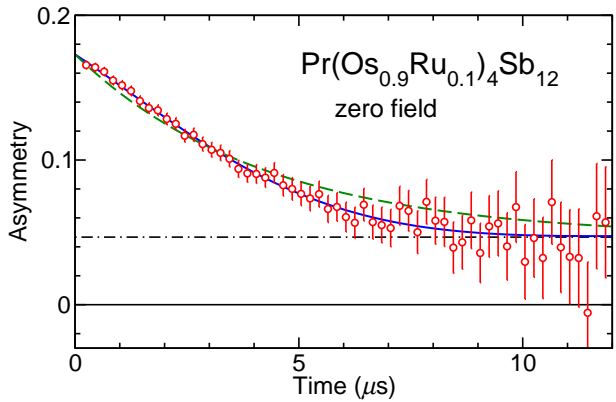}
 \caption{(color online) Zero-field positron count rate asymmetry (muon spin relaxation function) in Pr(Os$_{0.9}$Ru$_{0.1}$)$_{4}$Sb$_{12}$, $T = 1.31$~K\@. Solid (blue) curve: fit of the exponentially damped Gaussian \mbox{K-T} function [Eq.~(\protect\ref{eq:ds})] to the data. Dashed (green) curve: best fit of a simple exponential to the data. Dash-dot line: time-independent signal from muons that stopped outside the sample.}
 \label{fig:ZF-asy}
 \end{center}
\end{figure}

Figure~\ref{fig:ZF-asy} shows the ZF-$\mu$SR positron count rate asymmetry\cite{Sche85,Brew94}~$AG(t)$, where $A$ is the initial asymmetry and $G(t)$ is the muon spin polarization (initially $\sim$100\%), in Pr(Os$_{0.9}$Ru$_{0.1}$)$_{4}$Sb$_{12}$.
The time-independent offset in Fig.~\ref{fig:ZF-asy} comes from nonrelaxing muons that stopped in the silver cold finger or elsewhere in the cryostat (i.e., did not stop in either the sample or the GaAs backing).

The zero-field static Gaussian \mbox{K-T} relaxation function\cite{KuTo67,HUIN79}
\begin{equation}
 \label{eq:KT}
 G_{z}^{\mathrm{KT}}(\Delta,t) = \frac{1}{3} + \frac{2}{3}(1-\Delta^{2}t^{2})\, \exp(-{\textstyle\frac{1}{2}}\Delta^{2}t^{2})
\end{equation}
describes muon spin relaxation by randomly-oriented quasistatic muon local fields, with Cartesian components that vary according to a Gaussian distribution with a zero mean value and an rms width~$\Delta/\gamma_\mu$ ($\gamma_{\mu}$ is the muon gyromagnetic ratio). The ``damped Gaussian \mbox{K-T}'' function
\begin{equation}
 \label{eq:ds}
 G(t) = \exp(-\Lambda t)\,G_{z}^{\mathrm{KT}}(\Delta,t) \,,
\end{equation}
where $\Lambda$ is the damping rate, was fit to the experimental data (solid curve in Fig.~\ref{fig:ZF-asy}). Although the data appear to decay roughly exponentially, it can be seen that the best fit to a simple exponential (dashed curve in Fig.~\ref{fig:ZF-asy}) is not as good as the fit to Eq.~(\ref{eq:ds}).

Before analyzing the relaxation data further, we discuss the choice of Eq.~(\ref{eq:ds}) as a fitting function. In ZF- and LF-$\mu$SR, fluctuations of $\mathbf{H}_{\mathrm{loc}}$ are often treated using the ``dynamic \mbox{K-T}'' relaxation function,\cite{HUIN79} in which $\mathbf{H}_{\mathrm{loc}}$ fluctuates {\em as a whole\/} with a single correlation time. Such a procedure is not appropriate in the present experiments, however. A quasistatic contribution to $\mathbf{H}_{\mathrm{loc}}$, due mainly to Sb nuclear dipolar fields, is also present, so that $\mathbf{H}_{\mathrm{loc}}$ is the sum of quasistatic and fluctuating components. \cite{ATKS03}

\begin{figure}[ht]
 \begin{center}
 \includegraphics*[width=0.45\textwidth]{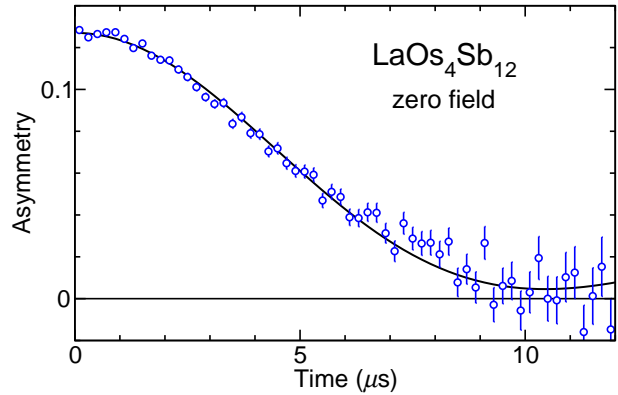}
 \caption{(color online) Zero-field muon spin polarization decay function in LaOs$_{4}$Sb$_{12}$, $T = 0.91$~K\@. Data from Ref.~\protect\onlinecite{AHSO05}. The signal from muons that do not stop in the sample has been subtracted. There is no significant exponential damping.}
 \label{fig:ZF-La}
 \end{center}
\end{figure}

Figure~\ref{fig:ZF-La} shows ZF-$\mu$SR data obtained from the isostructural compound~LaOs$_{4}$Sb$_{12}$ (here the time-independent signal has been subtracted), in which there are no $4f$ electrons. \cite{AHSO05}
The best fit of Eq.~(\ref{eq:ds}) to these data yields a negligible exponential damping rate, indicating that the damping arises from the presence of Pr ions. The values of the \mbox{K-T} relaxation rate~$
$ in the normal state differ by only $\sim$15\% between PrOs$_{4}$Sb$_{12}$ [$\Delta = 0.143(5)~\mu\mathrm{s^{-1}}$] and LaOs$_{4}$Sb$_{12}$ [$\Delta = 0.167(5)~\mu\mathrm{s^{-1}}$]. In the latter compound the nuclear dipolar field is the only mechanism for the static \mbox{K-T} term. \cite{AHSO05,Larelax}

This near equality is additional evidence that $\Delta$ is due to nuclear dipolar fields in PrOs$_4$Sb$_{12}$. Furthermore, significantly better fits to zero-field data are obtained using the damped static \mbox{K-T} function (reduced $\chi^2$ typically $\sim$1.1) than the dynamic \mbox{K-T} function (reduced $\chi^2$ typically $\sim$1.4). We therefore use damped static relaxation of the form of Eq.~(\ref{eq:ds}) to model the situation where muon spin states are split into Zeeman levels by the quasistatic component of $\mathbf{H}_L + \mathbf{H}_{\mathrm{loc}}$, and the fluctuating component of $\mathbf{H}_{\mathrm{loc}}$ induces transitions between these Zeeman levels.

\begin{figure}[ht]
 \begin{center}
 \includegraphics*[width=0.45\textwidth]{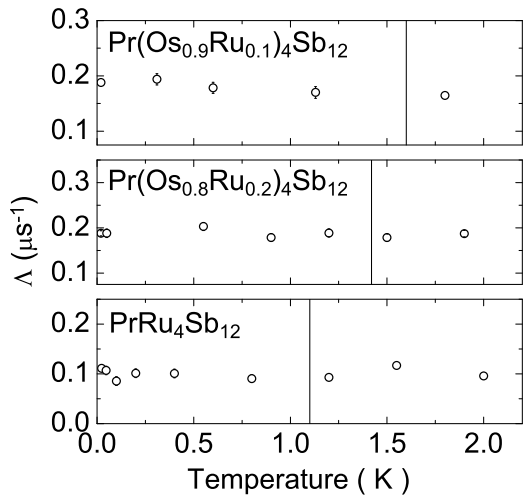}
 \caption{Zero-field exponential damping rate~$\Lambda$ vs temperature
 in Pr(Os$_{1-x}$Ru$_x$)$_4$Sb$_{12}$, $x = 0.1$, 0.2, and 1.0.
 The vertical lines indicate the superconducting transition temperatures for each alloy.}
 \label{fig:Zf-Rel}
 \end{center}
\end{figure}

In Fig.~\ref{fig:Zf-Rel} the damping rate~$\Lambda$ in zero field for Pr(Os$_{1-x}$Ru$_x$)$_4$Sb$_{12}$, $x = 0.1$, 0.2, and 1.0, is plotted vs temperature~$T$ for $0.02~{\mathrm{K}} \le T \le 2~{\mathrm{K}}$.
As in PrOs$_{4}$Sb$_{12}$,\cite{ATKS03} little temperature dependence is observed in this temperature range; this is also the case for Pr$_{1-y}$La$_{y}$Os$_{4}$Sb$_{12}$ alloys (data not shown). This weak temperature dependence strongly suggests that $\Lambda$ is due to nuclear magnetism rather than electronic spin fluctuations: the latter would be expected to show significant temperature dependence, especially below $T_c$, whereas fluctuations arising from nuclear spin-spin interactions are temperature-independent except at very low temperatures. \cite{Abra61} The dynamic relaxation is unlikely to arise from fluctuations of $^{121}$Sb and $^{123}$Sb nuclear spins, which should result in quasistatic contributions to $\mathbf{H}_{\mathrm{loc}}$ because their spin-spin and spin-lattice relaxation times are relatively long. \cite{KYIK03}

A candidate mechanism for dynamic muon spin relaxation is the hyperfine-enhanced $^{141}$Pr nuclear spin system with hyperfine-enhanced effective nuclear moments,\cite{Blea73}, described briefly in Sec.~\ref{sect:Intro}, which as discussed below in Sec.~\ref{sect:LF} leads to dynamic muon spin relaxation rates in qualitative agreement with experiment. $^{141}$Pr nuclear spin fluctuations due to hyperfine-enhanced $^{141}$Pr-$^{141}$Pr spin-spin interactions can be rapid enough to cause dynamic muon spin relaxation, since both dipolar and exchange-mediated contributions to these interactions are increased by hyperfine enhancement. The observed nuclear Schottky anomaly in the low-temperature specific heat of PrOs$_{4}$Sb$_{12}$,\cite{ANOS02} is similarly enhanced. Hyperfine-enhanced relaxation is independent of temperature for $T \ll \Delta E_{\mathrm{CEF}}/k_B$,\cite{MHNC00} which is roughly satisfied in the present experiments.

\subsection{\label{sect:LF}Longitudinal-field muon spin relaxation}
We first review the effect of longitudinal applied field~$H_L$ on LF-$\mu$SR measurements. \cite{KuTo67,HUIN79} In the case where $\mathbf{H}_{\mathrm{loc}}$ is quasistatic, for $H_L \gg \Delta/\gamma_\mu$ the (quasistatic) resultant field~$\mathbf{H}_{\mathrm{loc}} + \mathbf{H}_L$ is nearly parallel to the initial muon spin direction. Then the precession that causes quasistatic relaxation is reduced in amplitude, and the muon spin polarization is nearly time-independent. For weaker fields~$H_L \approx \Delta/\gamma_\mu$, the muon polarization at long times increases with increasing $H_L$. This phenomenon, called ``decoupling,'' does not occur for dynamic relaxation unless the latter is quenched for $H_L \approx \Delta/\gamma_\mu$; this is unusual because nuclear dipolar fields are small, of the order of a few Oe, and applied fields this small seldom affect dynamic relaxation mechanisms. Thus LF-$\mu$SR measurements help to determine whether the observed ZF relaxation is due to static or dynamic contributions to $\mathbf{H}_{\mathrm{loc}}$. \cite{HUIN79} In favorable cases the dependence of the damping rate~$\Lambda$ on $H_L$ yields statistical properties (rms amplitude, correlation time) of the fluctuating field. \cite{LoTs68,HUIN79}

\begin{figure}[ht]
 \begin{center}
 \includegraphics*[width=0.45\textwidth]{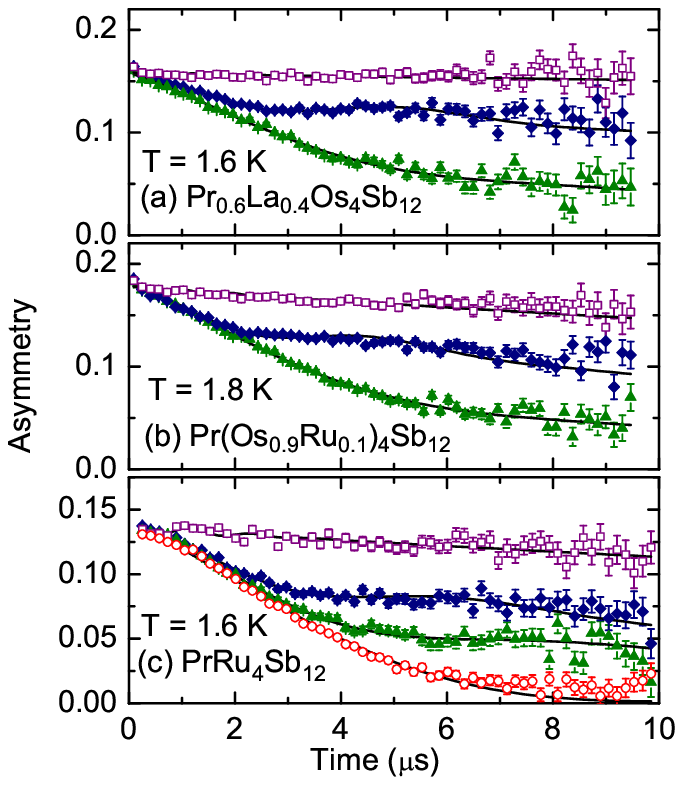}
 \caption{(color online). Representative ZF- and LF-$\mu$SR spin polarization decay functions. (a)~Pr$_{0.6}$La$_{0.4}$Os$_{4}$Sb$_{12}$. Triangles: $H_L$ = 5.9~Oe. Diamonds: $H_L$ = 16.1~Oe. Squares: $H_L$ = 121~Oe. (b)~Pr(Os$_{0.9}$Ru$_{0.1}$)$_{4}$Sb$_{12}$. Triangles: $H_L$ = 5.3~Oe. Diamonds: $H_L$ = 16.0~Oe. Squares: $H_L$ = 75.6~Oe. (c)~PrRu$_{4}$Sb$_{12}$. Circles: $H_L$ = 0~Oe. Triangles: $H_L$ = 6.3~Oe. Diamonds: $H_L$ = 10~Oe. Squares: $H_L$ = 63~Oe. Curves: fits to the damped static \mbox{K-T} function in longitudinal field [Eq.~(\protect\ref{eq:ds})] (see text).}
 \label{fig:LF-asy}
 \end{center}
\end{figure}

LF-$\mu$SR experiments were performed in weak longitudinal fields~$H_L$ in the normal state just above $T_c$. Representative LF-$\mu$SR spectra are shown in Fig.~\ref{fig:LF-asy}.
The data exhibit the late-time field dependence that is a characteristic feature of decoupling, together with overall damping that is stronger at lower fields. A ``damped static longitudinal \mbox{K-T} function'' appropriate to nonzero $H_L$, of the form of Eq.~(\ref{eq:ds}) with $G_{z}^{\mathrm{KT}}(\Delta,t)$ now the static Gaussian \mbox{K-T} function in nonzero applied longitudinal field,\cite{HUIN79} was therefore fit to the data. ``Global'' fits to all the field-dependent data at a given temperature were carried out, with $\Delta$ taken to be independent of field but varied for best fit. The field dependence of $\Lambda$ was obtained under this condition.

\begin{figure*}[ht]
 \begin{center}
 \includegraphics*[width=0.9\textwidth]{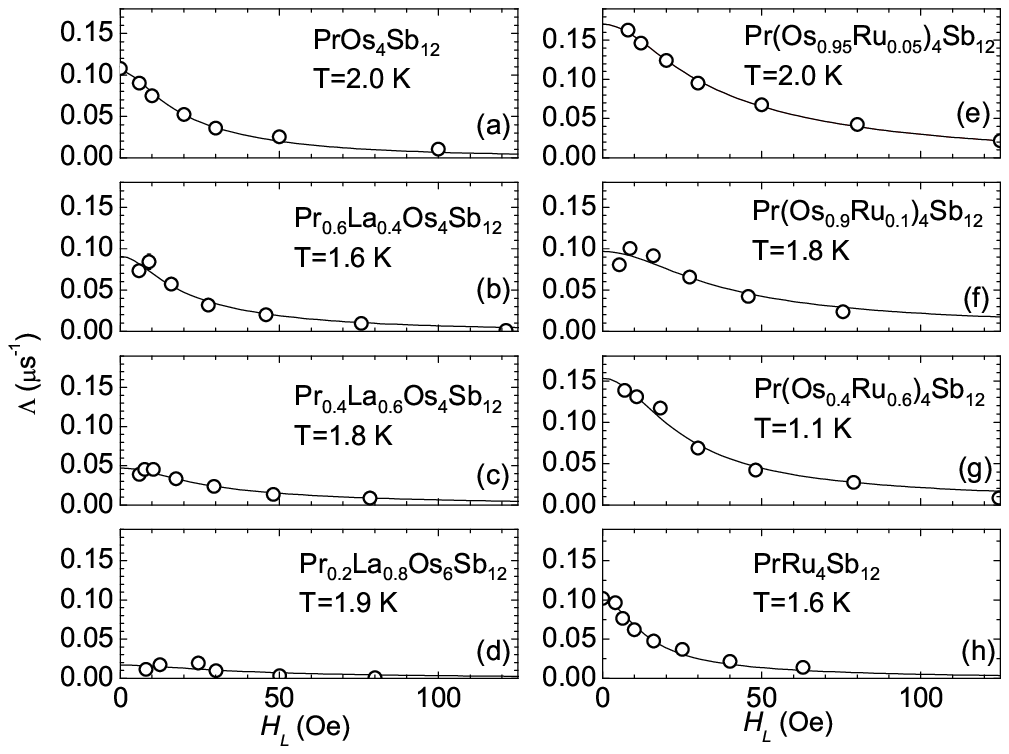}
 \caption{Exponential damping rate~$\Lambda$ vs longitudinal applied field~$H_L$ in PrOs$_{4}$Sb$_{12}$ and its alloys. (a):~PrOs$_{4}$Sb$_{12}$ (data from Ref.~\protect\onlinecite{ATKS03}). (b)-(d):~Pr$_{1-y}$La$_{y}$Os$_{4}$Sb$_{12}$, $y = 0.4$, 0.6, and 0.8. (e)-(h):~Pr(Os$_{1-x}$Ru$_{x}$)$_{4}$Sb$_{12}$, $x = 0.05$, 0.1, 0.6, and 1. Curves: fits to Eq.~(\protect\ref{eq:Red}) (Ref.~\protect\onlinecite{HUIN79}).}
 \label{fig:LF-Lam}
 \end{center}
\end{figure*}

Figure~\ref{fig:LF-Lam} shows these field dependencies for several alloys.
The fit values of $\Delta$ are given in Table~\ref{table-LF}, which also shows parameters derived from the field dependence of $\Lambda$ as discussed below in Sec.~\ref{sec:LT}.
\begin{table*}[ht]
\caption{Parameters from analysis of LF-$\mu$SR data in PrOs$_{4}$Sb$_{12}$, Pr$_{1-y}$La$_{y}$Os$_{4}$Sb$_{12}$ ($y=0.4$, 0.6, and 0.8), Pr(Os$_{1-x}$Ru$_{x}$)$_{4}$Sb$_{12}$ ($x=0.05$, 0.1, and 0.6), and PrRu$_{4}$Sb$_{12}$. $\Delta$: quasistatic \mbox{K-T} relaxation rate obtained from longitudinal-field fits to Eq.~(\protect\ref{eq:ds}) (see text) except for PrOs$_{4}$Sb$_{12}$. $K_{\chi}$: hyperfine enhancement factor (Knight shift) calculated from observed low-temperature molar susceptibility (Refs.~\protect\onlinecite{Blea73} and \protect\onlinecite{MHNC00}) and Eq.~(\protect\ref{eq:hhf}). $\sigma_\mathrm{VV}$: $\mu^+$-$^{141}$Pr Van Vleck relaxation rate (used as measure of $\mu^+$-$^{141}$Pr coupling strength). $\tau_{c}$: correlation time of $^{141}$Pr nuclear spin fluctuations. Experimental values (exp) obtained from fits of Eq.~(\protect\ref{eq:Red}) to longitudinal field dependence of damping rate $\Lambda$ (Fig.~\protect\ref{fig:LF-Lam}). Calculated values (calc) obtained from lattice sums assuming dipolar coupling (see text).} \label{table-LF}
\begin{ruledtabular}
\begin{tabular}{lcccccc}
 Alloy & $\Delta$ ($\mu\mathrm{s}^{-1}$) & $K_{\chi}$ & \multicolumn{2}{c}{$\sigma_\mathrm{VV}$ ($\mu\mathrm{s}^{-1}$)} & \multicolumn{2}{c}{$\tau_{c}$ ($\mu\mathrm{s}$)} \\
 &  &  & exp & calc & exp & calc \\
 \hline
 PrOs$_{4}$Sb$_{12}$\footnotemark[1] 
 & 0.14\footnotemark[2] &  19.7\footnotemark[3]  & 0.26(1) & 0.34 & 0.31(2) & 1.31\\
 Pr$_{0.6}$La$_{0.4}$Os$_{4}$Sb$_{12}$ & 0.21 & 16.3  & 0.24(1) & 0.22 & 0.32(6) & 2.41 \\
 Pr$_{0.4}$La$_{0.6}$Os$_{4}$Sb$_{12}$ & 0.21 & 14.6  & 0.20(1) & 0.16 & 0.24(4) & 3.59 \\
 Pr$_{0.2}$La$_{0.8}$Os$_{4}$Sb$_{12}$ & 0.21 & 14.7 & 0.14(3) & 0.12 & 0.2(1) &  5.02 \\
 Pr(Os$_{0.95}$Ru$_{0.05}$)$_{4}$Sb$_{12}$ & 0.25 &  15.9\footnotemark[4]  & 0.43(1) & 0.28 &  0.19(1) & 1.96 \\
 Pr(Os$_{0.9}$Ru$_{0.1}$)$_{4}$Sb$_{12}$ & 0.24 & 12.2\footnotemark[3]  & 0.33(3) & 0.22 & 0.18(4) & 3.22 \\
 Pr(Os$_{0.4}$Ru$_{0.6}$)$_{4}$Sb$_{12}$ & 0.38 &  7.51\footnotemark[3]  & 0.31(1) & 0.14 & 0.31(2) & 7.66\\
 PrRu$_{4}$Sb$_{12}$ & 0.22 &  6.76\footnotemark[3]  & 0.18(1) & 0.13 & 0.60(6) & 9.19\\
\end{tabular}
\end{ruledtabular}
\footnotetext[1]{LF-$\mu$SR data from Ref.~\protect\onlinecite{ATKS03}.}
\footnotetext[2]{From ZF-$\mu$SR data of Ref.~\protect\onlinecite{ATKS03}.}
\footnotetext[3]{Susceptibility data from Ref.~\protect\onlinecite{FDHB04}.}
\footnotetext[4]{Estimated from susceptibility data of Ref.~\protect\onlinecite{FDHB04}.}
\end{table*}
The values of $\Delta$ obtained in this way are larger than those obtained from fits to ZF-$\mu$SR data, and moreover $\Delta$ exhibits an increase with increasing $H_L$ (not shown) if allowed to vary with field. This behavior indicates that the data depart somewhat from the form of Eq.~(\ref{eq:ds}), and we consider the fit values of $\Delta$ and $\Lambda$ to be of qualitative significance only.

If the exponential damping were due to a (Lorentzian) distribution of static contributions to $\mathbf{H}_{\mathrm{loc}}$, the observed ZF value of $\Lambda$ leads to an estimate~$\Lambda/\gamma_\mu \sim 1$~Oe for the spread of the local fields~$H_{\mathrm{loc}}$. Then an applied longitudinal field of order 10~Oe should nearly decouple $H_{\mathrm{loc}}$ and there should be almost no damping. But it can be seen from Fig.~\ref{fig:LF-Lam} that in general $\Lambda$ is reduced only slightly for $H_L$ = 10~Oe. Thus the exponential damping is not decoupled, which is evidence that $\Lambda$ is dynamic rather than static in origin. \cite{ATKS03} The observed decoupling in Fig.~\ref{fig:LF-asy} is associated solely with the behavior of $G_{z}^{\mathrm{KT}}(\Delta,t)$, hence with quasistatic relaxation by Sb nuclear dipolar fields.

For any distribution of quasistatic local fields the ZF-$\mu$SR asymmetry at long times is expected to approach 1/3 of its initial value~$A$. \cite{HUIN79} In Pr-based samples, however, the zero-field value of the asymmetry at long times is typically much lower than $A/3$. For PrRu$_4$Sb$_{12}$ this can be seen in Fig.~\ref{fig:LF-asy}(c) (circles) (see also Fig.~2 of Ref.~\onlinecite{AHPG05}). Like the absence of decoupling, this behavior is evidence that the exponential damping is dynamic.

\section{Discussion} \label{sec:discussion}
\subsection{Muon relaxation by fluctuating $^{\bm{141}}
$Pr nuclear spins} \label{sec:LT}
Hayano {\em et al.\/}\cite{HUIN79} calculated the LF-$\mu$SR and TF-$\mu$SR relaxation rates due to dipole-coupled nuclear spins based on the theory of magnetic resonance absorption formulated by Kubo and Tomita. \cite{KuTo54} In their calculation spin dynamics arise from muon diffusion in a lattice of quasistatic nuclear spins, but the treatment applies equally to the case where the muon is stationary and it is the nuclear spins that are fluctuating; Lowe and Tse\cite{LoTs68} carried out an equivalent calculation for relaxation of nuclear spins, with essentially the same result. We therefore apply the results of Hayano {\em et al.}\ to the present experiments, after modification to include hyperfine enhancement of the $^{141}$Pr nuclear spins. This is accomplished by replacing the bare $^{141}$Pr gyromagnetic ratio~$^{141}\gamma$ by the enhanced value~$^{141}\gamma(1+K)$, where $K$ is the hyperfine enhancement factor (i.e., the $^{141}$Pr Knight shift),\cite{Blea73,MHNC00} assumed isotropic for the tetrahedral\cite{THY01} Pr site. The damping rate~$\Lambda$ is
\begin{eqnarray}
 \label{eq:Red}
 \Lambda
  & = & \frac{\sigma_{\mathrm{VV}}^{2}}{2}\left\{\frac{3\tau_{c}}{1 + (\gamma_{\mu}H_L\tau_{c})^{2}}\right.\nonumber \\
  &  & \qquad\quad +\ \frac{\tau_{c}}{1 + [\gamma_{\mu} - ^{141}\gamma(1+K)]^{2}H_L^{2}\tau_{c}^{2}}\nonumber \\
  &  & \qquad\quad + \left.\frac{6\tau_{c}}{1+[\gamma_{\mu} + ^{141}\gamma(1+K)]^{2}H_L^{2}\tau_{c}^{2}}\right\} \,,
\end{eqnarray}
where $\sigma_{\mathrm{VV}}$ is the high-field muon Van Vleck relaxation rate\cite{HUIN79} due to $^{141}$Pr dipolar fields (used here as a measure of the rms amplitude of these fields), $\tau_c$ is the correlation time of the $^{141}$Pr spin fluctuations, and a powder average has been taken. The rapid-fluctuation limit~$\sigma_{\mathrm{VV}}\tau_c \ll 1$ is assumed, since otherwise an exponential damping function would not be expected. \cite{HUIN79} This relation includes the contributions to $\Lambda$ of both longitudinal $^{141}$Pr spin fluctuations,\cite{LoTs68} which are assumed to have a common correlation time~$\tau_c$. The field dependence of $\Lambda$ arises from the fact that the muon spin relaxation rate is proportional to the fluctuation noise power at the muon Zeeman frequency~$\gamma_\mu H_L$; for weakly-coupled nuclear spins the dipolar fluctuation spectrum consists of broadened peaks centered at 0, $^{141}\gamma(1 + K)H_L$, and $-[^{141}\gamma(1 + K)H_L]$,\cite{Abra61,LoTs68} corresponding to each of the terms of Eq.~(\ref{eq:Red}).

From Fig.~\ref{fig:LF-Lam} it can be seen that the field dependence of $\Lambda$ is generally well fit by Eq.~(\ref{eq:Red}). We find, however, that the form of this relation does not determine $K$ well compared to the other fitting parameters. We therefore obtain $K$ independently from the observed low-temperature molar susceptibility $\chi_{\mathrm{mol}}$ of the Pr ions,\cite{FDHB04,RKA06,AYO06} and fix it in the fitting. This is done using the relation\cite{Blea73,AnDa77,MHNC00}
\begin{equation}
K = K_\chi = a_{\mathrm{hf}}\chi_{\mathrm{mol}} \,,
\label{eq:hhf}
\end{equation}
where $a_{\mathrm{hf}} = 187.7$ mole emu$^{-1}$ is the Pr atomic hyperfine coupling constant. The calculated values of $K_\chi$ are given in Table~\ref{table-LF}, together with the experimental values of $\sigma_{\mathrm{VV}}$ and $\tau_{c}$ obtained from the fits.

\subsection{Comparison of data to hyperfine-enhancement model} \label{sec:comparison}
Next we compare the experimental values of $\sigma_{\mathrm{VV}}$ and $\tau_c$ to those expected from the hyperfine-enhancement scenario, assuming dipolar couplings between all spins. We make the ansatz that the $^{141}$Pr spin fluctuations are due to spin-spin interactions within the $^{141}$Pr spin system; in analogy with the discussion of muon spin relaxation in Sec.~\ref{sec:ZF}, one would expect significant $^{141}$Pr spin-lattice relaxation by 4$f$ electronic spin fluctuations to result in considerable temperature dependence of $\Lambda$ contrary to experiment.

We first calculate the ``unenhanced'' (i.e., using the bare $^{141}$Pr gyromagnetic ratio) powder-average $\mu^+$-$^{141}$Pr Van Vleck relaxation rate $\sigma_{\mathrm{VV}}^{0}$ from a standard lattice sum~\cite{HUIN79,Abra61} for the candidate~$\left( 0, \frac{1}{2}, 0.15\right)$ muon site\cite{ATKS03} (12$e$ in Wyckoff notation) in the filled-skutterudite structure (space group $Im\bar{3}$). For dipolar $^{141}$Pr-$^{141}$Pr coupling the correlation time $\tau_{c}^{0}$ in the absence of hyperfine enhancement is estimated by the inverse of the unenhanced like-spin $^{141}$Pr-$^{141}$Pr Van Vleck rate, which is obtained from a lattice sum similar to that for $\sigma_{\mathrm{VV}}^0$. \cite{Abra61} The hyperfine-enhanced values of these quantities are then\cite{Blea73,MHNC00}
\begin{equation}
 \label{eq:sigma}
 \sigma_{\mathrm{VV}}=\sigma_{\mathrm{VV}}^{0} (1+K)
\end{equation}
and
\begin{equation}
 \label{eq:tau}
 \tau_{c}=\tau_{c}^{0}/(1+K)^{2} \,;
\end{equation}
in both cases the dependence on $K$ comes solely from the hyperfine enhancement of $^{141}\gamma$ (Refs.~\onlinecite{Blea73} and \onlinecite{MHNC00}).

The calculated values of $\sigma_{\mathrm{VV}}$ and $\tau_{c}$ are given in Table~\ref{table-LF}. For the La-doped alloys ensemble averages of the lattice sums have been taken over random $^{141}$Pr site locations. \cite{Abra61} The experimental and calculated values of $\sigma_{\mathrm{VV}}$ are in rough agreement, but the experimental values of $\tau_c$ are systematically smaller than the calculated values, sometimes by more than an order of magnitude. Correspondingly, the inequality~$\sigma_{\mathrm{VV}}\tau_c \ll 1$ required for our analysis is satisfied by the experimental values of $\sigma_{\mathrm{VV}}$ and $\tau_c$ but usually not by the calculated values.

The discrepancy between experimental and calculated values of $\tau_c$ is most likely an indication that the assumption of purely dipolar $^{141}$Pr-$^{141}$Pr interactions is not valid. Indirect RKKY-like interactions mediated by the Pr$^{3+}$ intra-ionic exchange interaction\cite{Blea73} may decrease $\tau_c$ considerably since they are also hyperfine-enhanced, but are difficult to estimate in PrOs$_4$Sb$_{12}$. A brief discussion of limits on the exchange-mediated $^{141}$Pr-$^{141}$Pr interaction constant~$^{141}{\cal J}_{\mathrm{ex}}$ is given in the Appendix, which concludes that indirect interactions would account for the experimental values of $\tau_c$ (Table~\ref{table-LF}) with only a modest Pr$^{3+}$ exchange coupling.

We note that $\tau_c$ is also the $^{141}$Pr NMR signal lifetime (spin-echo decay time)~$T_2$, so that $^{141}$Pr NMR experiments would provide an independent measure of this quantity. Unfortunately the values of $\tau_c$ from Table~\ref{table-LF} (${\lesssim}1\ \mu$s) are too short for the NMR signal to be observable using current spectrometer technology. Moreover, a search for the $^{141}$Pr resonance would be difficult because of uncertainty in the hyperfine enhancement factor~$K$\@. The mere observation of a $^{141}$Pr NMR signal would therefore significantly modify the conclusions of this paper, a fact which might motivate such a search.

An exchange-mediated $^{141}$Pr-$\mu^+$ interaction, using a Fermi contact interaction between the mediating electrons and the muon, could also be present. A scalar interaction $\propto  \mathbf{I}{\,\cdot\,}\mathbf{S}_\mu$, where $\mathbf{I}$ and $\mathbf{S}_\mu$ are $^{141}$Pr nuclear and muon spin operators, respectively, leads to a muon spin relaxation rate of the form of Eq.~(\ref{eq:Red}) but with only the second term in the brackets. \cite{DKCG06} Unfortunately the relative strengths of dipolar and exchange-mediated interactions cannot be determined accurately from fits of an appropriately generalized version of Eq.~(\ref{eq:Red}) to the data. Exchange-mediated interactions would, however, be unlikely to dominate (hyperfine-enhanced) $\mu^+$-$^{141}$Pr dipolar interactions, since in $f$-electron metals the magnitudes of dipolar and electron-mediated interactions between local electronic moments and muon spins are usually comparable. \cite{Amat97} The rough agreement between experimental and calculated dipolar values of $\sigma_\mathbf{VV}$ (Table~\ref{table-LF}) is consistent with this observation.

Other sources of uncertainty in the comparison between experimental and calculated parameters in Table~\ref{table-LF} include (1)~the fact that the low-temperature susceptibility may not be entirely due to the Van Vleck mechanism, leading to error in the calculation of $K_\chi$, and (2)~the fact that the calculation of $\sigma_{\mathrm{VV}}$ depends on the assumed muon stopping site in PrOs$_4$Sb$_{12}$. The stopping site has not been determined definitively; indeed, high-field TF-$\mu$SR\cite{HSKO07} suggests that there may be more than one muon site. Given these caveats, together with the qualitative nature of parameters derived from fitting the $\mu$SR data (Sec.~\ref{sect:LF}), the agreement between the experimental results and the hypothesis of hyperfine-enhanced $^{141}$Pr nuclear magnetism can be regarded as satisfactory.

\begin{figure}[ht]
 \begin{center}
 \includegraphics*[width=0.45\textwidth]{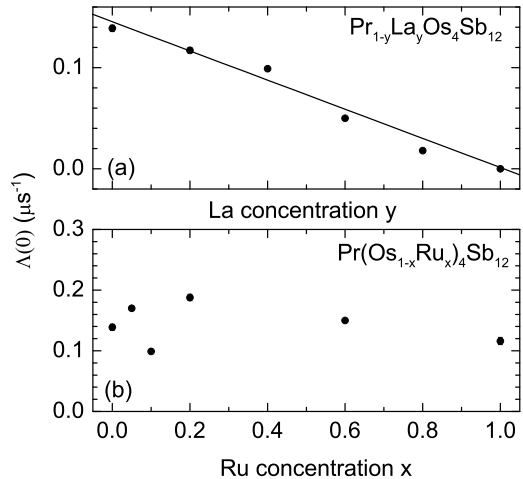}
 \caption{Concentration dependence of zero-field low-temperature exponential damping rate $\Lambda(0)$ in (a): Pr$_{y}$La$_{1-y}$Os$_{4}$Sb$_{12}$, (b): Pr(Os$_{x}$Ru$_{1-x}$)$_{4}$Sb$_{12}$.}
 \label{fig:xy}
 \end{center}
\end{figure}

Figure~\ref{fig:xy} gives the Ru and La concentration dependencies of the zero-field damping rate $\Lambda(0)$ just above $T_c$.
We see that $\Lambda(0)$ decreases as the Pr sublattice is diluted with La ions, as expected if the dynamic relaxation is due to $^{141}$Pr nuclear magnetism. In Pr(Os$_{1-x}$Ru$_x$)$_4$Sb$_{12}$, where the Pr sublattice is not diluted, it can be seen in Fig.~\ref{fig:xy}(b) that $\Lambda(0)$ generally increases by $\sim$50\% as $x$ is increased from zero, and decreases again as $x \rightarrow 1$. For dipolar coupling $\Lambda(0) = 5\sigma_{\mathrm{VV}}^{2}\tau_{c}$ [cf.\ Eq.~(\ref{eq:Red})], so that the $K$ dependencies of $\sigma_{\mathrm{VV}}$ and $\tau_{c}$ [Eqs.~(\ref{eq:sigma}) and (\ref{eq:tau})] cancel and there should be no dependence of $\Lambda(0)$ on Ru concentration. But this cancellation would not necessarily hold for an exchange-mediated interaction, which we have seen is necessary to understand the short experimental values of $\tau_c$. The Ru concentration dependence is not well understood, but at least two mechanisms may be involved: (1)~the concentration dependence of $\Delta E_\mathrm{CEF}$, which affects $\chi_\mathrm{mol}$ and hence $K_\chi$ [Eq.~(\ref{eq:hhf})], and (2)~the (unknown) concentration dependence of $^{141}{\cal J}_{\mathrm{ex}}$ across the alloy series.

\section{Conclusions} \label{sec:concl}
ZF- and LF-$\mu$SR measurements have been carried out in the filled skutterudite alloys~Pr(Os$_{1-x}$Ru$_x$)$_{4}$Sb$_{12}$, $x = 0.05$, 0.1, 0.2, 0.6, and 1.0, and (Pr$_{1-y}$La$_y$)Os$_{4}$Sb$_{12}$, $y = 0.2$, 0.4, 0.6, and 0.8, to clarify the origin of the dynamic damping of the muon spin relaxation observed in these alloys. At low temperatures, LF-$\mu$SR experiments in both alloy series indicate that static local field distributions and dynamic fluctuations are both involved in muon spin relaxation. The temperature and concentration dependencies of the muon damping rate $\Lambda$ suggest that it is due to hyperfine-enhanced $^{141}$Pr nuclear magnetism; the enhancement is responsible for the increased $^{141}$Pr nuclear spin-spin interaction strength and consequent rapid $^{141}$Pr spin fluctuations. Further evidence for this picture comes from the field dependence of $\Lambda$, which is in reasonable agreement with fits to the model of Hayano {\em et al.\/}\cite{HUIN79} and calculated coupling strengths (Table~\ref{table-LF}) assuming hyperfine enhancement and exchange-mediated spin-spin coupling between $^{141}$Pr nuclei.

We conclude that hyperfine-enhanced $^{141}$Pr nuclear spin fluctuations account for the observed exponential damping of the muon spin relaxation function in PrOs$_4$Sb$_{12}$ and its alloys, and that electronic spin fluctuations (other than the Pr$^{3+}$ electronic response involved in hyperfine enhancement) are not directly involved in the muon spin relaxation.

\begin{acknowledgments}
We are grateful to D. Arseneau, B. Hitti, S.~R. Kreitz\-man, K. Nagamine, and K. Nishiyama for help during the experiments, to S.~K. Kim for help with sample preparation, and to W.~G. Clark and C.~M. Varma for valuable discussions. This work was supported by The U.S. NSF, grant nos.~0102293 and 0422674 (Riverside), 0203524 and 0604015 (Los Angeles), and 0335173 (San Diego), the U.S. DOE, contract DE-FG-02-04ER46105 (San Diego), and by the Grant-in-Aid for Scientific Research Priority Area ``Skutterudite'' No.~15072206 (Tokyo) of the Ministry of Education, Culture, Sports, Science and Technology, Japan. One of us (T.~Y.) wishes to acknowledge a JSPS Postdoctoral Fellowship for Research Abroad.
\end{acknowledgments}

\appendix* \section{Exchange-mediated $^{\bm{141}}$P\lowercase{r}-$^{\bm{141}}$P\lowercase{r} spin interaction}
Consider a $^{141}$Pr-$^{141}$Pr spin interaction Hamiltonian of the form
\begin{eqnarray}
^{141}{\cal H} & = &^{141}{\cal H}_\mathrm{dip} + ^{141}{\cal H}_\mathrm{ex} \nonumber \\
& = & ^{141}{\cal J}_{\mathrm{dip}} \left[3({\mathbf{I}_1{\,\cdot\,}\mathbf{r}_{12}})({\mathbf{I}_2{\,\cdot\,}\mathbf{r}_{12}})/r_{12}^2 - \mathbf{I}_1{\,\cdot\,}\mathbf{I}_2\right] \nonumber \\
& & \ +\ ^{141}{\cal J}_{\mathrm{ex}} \mathbf{I}_1{\,\cdot\,}\mathbf{I}_2 \,,
\end{eqnarray}
where $^{141}{\cal J}_{\mathrm{dip}} = [^{141}\gamma(1+K)\hbar]^2/r_{12}^3$ and $^{141}{\cal J}_{\mathrm{ex}}$ are the interaction constants for $^{141}$Pr-$^{141}$Pr dipolar and indirect exchange couplings, respectively, and $\mathbf{r}_{12}$ is the distance between $^{141}$Pr spins~$\mathbf{I}_1$ and $\mathbf{I}_2$. A scalar exchange interaction has been assumed for simplicity. Both $^{141}{\cal J}_{\mathrm{dip}}$ and $^{141}{\cal J}_{\mathrm{ex}}$ are hyperfine enhanced. In a simplified model (singlet ground and excited CEF-split states), \cite{Andr73,Blea73} the ratio~$^{141}{\cal J}_{\mathrm{ex}}/^{141}{\cal J}_{\mathrm{dip}}$ is given approximately by
\begin{equation}
^{141}{\cal J}_{\mathrm{ex}}/^{141}{\cal J}_{\mathrm{dip}} \approx {\cal J}_{\mathrm{ex}}^{\mathrm{el}}/{\cal J}_{\mathrm{dip}}^{\mathrm{el}} \,,
\label{eq:exchange}
\end{equation}
where ${\cal J}_{\mathrm{ex}}^{\mathrm{el}}$ and ${\cal J}_{\mathrm{dip}}^{\mathrm{el}}$ are the corresponding interaction constants for Pr$^{3+}$ interionic exchange and dipolar coupling.

To our knowledge ${\cal J}_{\mathrm{ex}}^{\mathrm{el}}$ has not been determined accurately in PrOs$_4$Sb$_{12}$ or its alloys. Fits to susceptibility data of CEF models using a molecular-field approximation to the exchange coupling~\cite{GORB06} yield a molecular field constant~$\lambda = 3.9$ mol/emu. This gives ${\cal J}_{\mathrm{ex}}^{\mathrm{el}}/k_B \approx N_\mathrm{A}\mu_\mathrm{eff}^2\lambda/k_Bz_\mathrm{eff} \approx 0.14$~K, where the $T=0$ Van Vleck effective moment~$\mu_\mathrm{eff} \approx 0.71\mu_\mathrm{B}$ has been assumed and $z_\mathrm{eff}$ is an effective number of nearest neighbors. This estimate should probably be considered a rough upper bound. It satisfies the criterion\cite{Blea63,Tram63}
\begin{equation}
{\cal J}_{\mathrm{ex}}^{\mathrm{el}} \ll \Delta E_{\mathrm{CEF}}
\label{eq:Jex}
\end{equation}
for the absence of exchange-induced magnetic ordering; as noted previously, $\Delta E_{\mathrm{CEF}}/k_B \approx 7$~K. There is no sign of magnetic ordering in PrOs$_4$Sb$_{12}$, and many of its electronic properties can be accounted for without invoking an exchange interaction. \cite{KINM03} The dipolar interaction~${\cal J}_{\mathrm{dip}}^{\mathrm{el}}$ can, however, be calculated: its maximum value is ${\cal J}_{\mathrm{dip}}^{\mathrm{el}}/k_B = (g\mu_\mathrm{B})^2/k_Br_\mathrm{nn}^3 \approx 8 \times 10^{-3}$~K, where $r_\mathrm{nn}$ is the nearest-neighbor Pr-Pr distance. A value of ${\cal J}_{\mathrm{ex}}^{\mathrm{el}}$ an order of magnitude larger than this would lead to $^{141}{\cal J}_{\mathrm{ex}}/^{141}{\cal J}_{\mathrm{dip}} \gg 1$ from Eq.~(\ref{eq:exchange}) without violating Eq.~(\ref{eq:Jex}), and could account for the experimental values of $\tau_c$ (Table~\ref{table-LF}).


\begin{thebibliography}{10}
\bibitem{quasistatic}
I.e., a field that varies slowly compared to the muon time scale of ${\sim}10\
  \mu\rm s$.
\bibitem{Larelax}
The slightly larger value of $\Delta$ in {LaOs$_4$Sb$_{12}$} may reflect the
  contribution of (unenhanced) $^{139}$La nuclear dipolar fields.
\end{thebibliography}
\end{document}